\documentclass[aps,prl,twocolumn,groupedaddress,showpacs,floatfix,superscriptaddress,longbibliography]{revtex4-2}
\usepackage[plainpages=false,pdfpagelabels,colorlinks=true,linkcolor=red,urlcolor=blue,citecolor=blue,pdftitle={Title},pdfauthor={},pdfdisplaydoctitle=true,pdfduplex=DuplexFlipLongEdge]{hyperref}
\usepackage{epsfig}
\usepackage{amsmath,amssymb}
\usepackage{physics}
\usepackage{graphicx}
\usepackage[dvipsnames,usenames]{xcolor}
\usepackage[normalem]{ulem}
\usepackage{todonotes}
\usepackage{soul} 

\begin{document}

\title{Magnetic, charge, and bond order in the two-dimensional {S}u-{S}chrieffer-{H}eeger-{H}olstein model}
\author{Max Casebolt}
\affiliation{Department of Physics, University of California, Davis, CA 95616,USA}
\author{Chunhan Feng}
\affiliation{Center for Computational Quantum Physics, Flatiron Institute, 162 5th Avenue, New York, NY, USA}
\affiliation{Department of Physics, University of California, Davis, CA 95616,USA}
\author{Richard T. Scalettar}
\affiliation{Department of Physics, University of California, Davis, CA 95616,USA}
\author{Steven Johnston}
\affiliation{Department of Physics and Astronomy, The University of Tennessee, Knoxville, TN 37996, USA}
\affiliation{Institute for Advanced Materials and Manufacturing, University of Tennessee, Knoxville, TN 37996, USA\looseness=-1}
\author{G. G. Batrouni}
\affiliation{Centre for Quantum Technologies, National University of
  Singapore, 2 Science Drive 3, 117542 Singapore}
\affiliation{Universit\'e C\^ote d'Azur, CNRS, Institut de Physique de Nice (INPHYNI), 06000 Nice, France}

\begin{abstract}
Most nonperturbative numerical studies of electron-phonon interactions focus on model Hamiltonians 
where the electrons interact with a phonon branch via a single type of microscopic mechanism. 
Two commonly explored couplings in this context are the Holstein and Su-Schrieffer-Heeger (SSH) 
interactions, which describe phonons modulating the on-site energy and intersite electron hopping, 
respectively. Many materials, however, have multiple phonon branches that can each interact with 
electronic degrees of freedom in different ways. We present here a determinant quantum Monte Carlo study of the half-filled two-dimensional (bond) SSH-Holstein Hamiltonian, where electrons
 couple to different phonon branches via either the Holstein or SSH mechanism. We map the model's 
phase diagram and determine the nature of the transitions between charge-density wave, bond order wave, and antiferromagnetic order.
\end{abstract} 

\maketitle



\noindent{\it Introduction} --- 
An electron interacting strongly with the lattice forms a composite quasiparticle known as a 
polaron~\cite{Devreese2009, Franchini2021}. Polarons can have large effective masses, which
 reflects the necessity of rearranging the lattice degrees of freedom as the electron 
moves~\cite{BonicPRB1999, Franchini2021, BerciuPRB2007, Devreese2009}. If two electrons 
are present, they can bind together in a bipolaron, which allows one electron to take advantage of the distortion produced by the other to occupy the same region of space. 
At half-filling and on a bipartite lattice, (bi)polarons tend to arrange themselves spatially
 into insulating charge density wave (CDW) or bond ordered wave (BOW) patterns, depending on
 the microscopic nature of the electron-phonon ($e$-ph) coupling~\cite{ScalettarPRB1989, EsterlisPRB2018, weber2018, Li20, feng2020, xing21, feng21, CohenStead22, gotz2022}. For the Holstein interaction, for 
example, bipolarons tend to freeze into an ordered ${\bf Q} = (\pi/a,\pi/a)$ lattice in two
 dimensions (where $a$ is the lattice constant) leading to a CDW insulating phase. Conversely, 
for the bond Su-Schrieffer-Heeger (SSH) interaction~\cite{sengupta03peierls} on a single-band
 lattice, antiferromagnetic (AFM) order can accompany BOW owing to a positive effective
 intersite exchange $J$ that appears when the phonons are integrated out~\cite{xing21, cai21, gotz2022}.
 However, optical SSH interactions on a multi-orbital Lieb or perovskite lattice can lead to a 
bipolaron lattice, much like the Holstein model for certain parameter regimes and
 filling values~\cite{Li20, CohenStead22}. 

The thermal and quantum phase transitions to these various ordered phases have mostly
 been studied via quantum Monte Carlo (QMC) for each interaction in isolation, i.e.~for 
either a Holstein coupling of the electron density to the phonon displacement~\cite{holstein1959, scalettar1989, marsiglio1990, vekic1992, batrouni2019, hohenadler2019, chen2019, zhang2019, bradley2021, feng2020i, feng2020, zhang2022, kvande2023} or an SSH coupling~\cite{Li20, xing21, cai21, feng21, nocera21, carbone21, gotz2022, zhang22, CohenStead22, xing2023, TanjaroonLy2023comparative}
 in which the phonon modulates the intersite hopping (kinetic energy), but not both.
 Yet in real materials with complex unit cells, several phonon branches can couple to 
the electrons. Moreover, the microscopic coupling mechanism to individual branches can 
be different, leading to opposing effects. The high-T$_c$ superconducting cuprates provide 
an interesting case in point. In these materials, the presence of a crystal field across 
the CuO$_2$ plane introduces an on-site coupling to the bond-buckling optical oxygen 
models \cite{Devereaux95, Johnston10}. At the same time, the bond-stretching Cu-O modes 
(the so-called half- and full-breathing modes) modulate the Cu-O hopping integral 
$t_{pd}$~\cite{Devereaux04, Johnston10}. The former coupling is naturally described by a 
Holstein-like interaction, while the latter is of the SSH type. When multiple coupling 
mechanisms are present, one naturally expects rich competition between the ordered phases 
created by the respective interactions. 

Despite this relevance of multi-branch models to real materials, combined SSH-Holstein 
(SSHH) models have previously been studied via QMC only in one dimension~\cite{hohenadler16}. 
In this case, the lower dimension precludes long-range order at nonzero temperature, resulting 
in ground state correlations with a power law decay with increasing distance. 
The key features of the 1D phase diagram are the presence of a metallic phase (spuriously 
absent in the pure 1D SSH model) when the Holstein coupling is introduced, and a competition 
between BOW and CDW correlations as the associated $e$-ph couplings are varied \cite{hohenadler16}. 
There is a direct and continuous transition between the states at strong coupling while a Luther-Emery metallic phase, in which the doublons and high kinetic energy bonds are disordered, 
intervenes at weak coupling where the interactions compete. 
The 1D SSHH model has also recently been studied in the context of designing topological 
analogs to magnetic bits~\cite{Xu2023}. 

\begin{figure}[t!]
\centering
\includegraphics[width=\columnwidth]{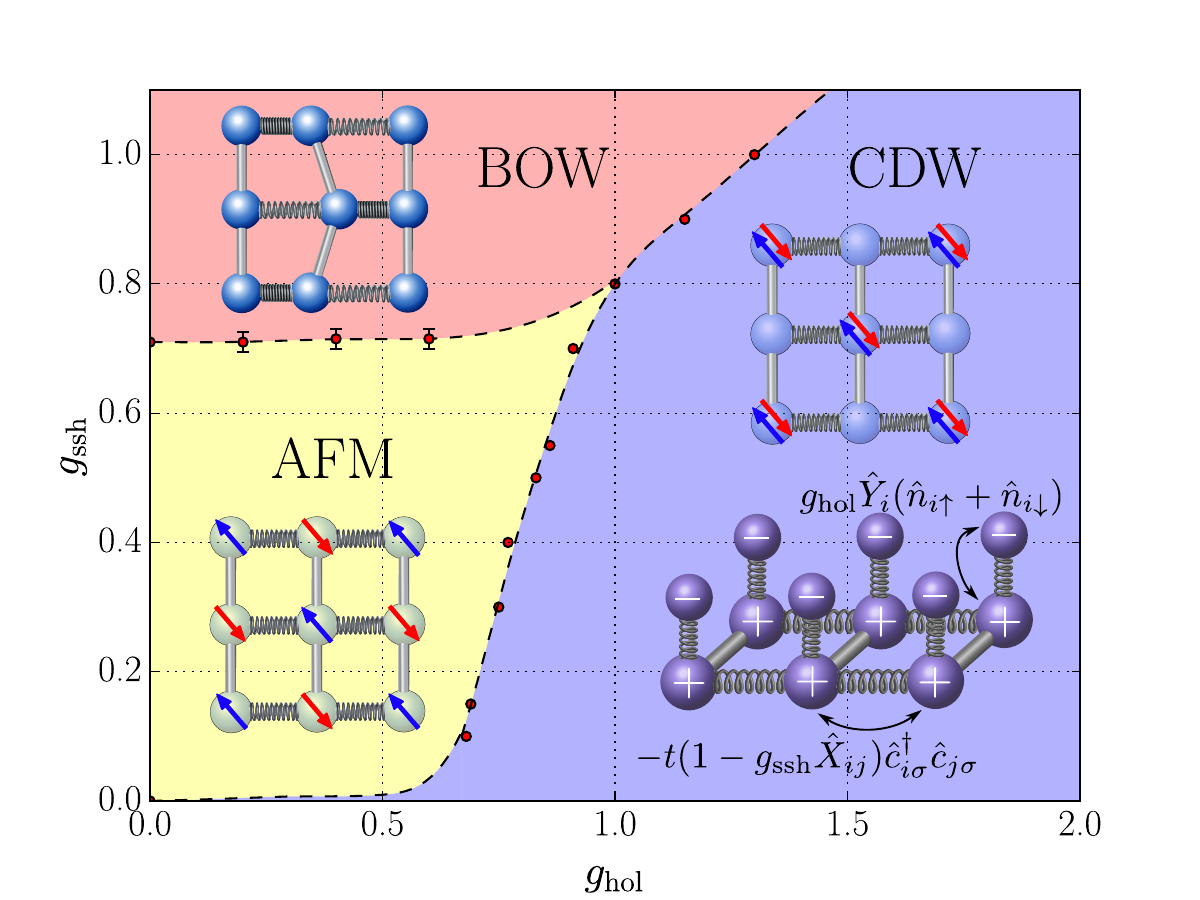} 
\caption{
The low-temperature ($T = t/16$) phase diagram of the SSH-Holstein model.
 CDW order is present at any $g_{\rm hol}$ for $g_{\rm ssh}=0$.
BOW dominates at large $g_{\rm ssh}$.
 For $g_{\rm ssh}\leq0.7$, antiferromagnetic order is present for low Holstein couplings
 in place of BOW.  Bottom right (in the blue zone): A sketch of the SSH-Holstein model. 
 The SSH interaction couples to the electron hopping, while the Holstein interaction 
couples to the electron density.  SSH phonons exist on lattice bonds while Holstein phonons exist on-site.
 }
\label{fig:Sallvsgholstfig1} 
\end{figure}

In this letter, we study the two-dimensional (2D) single-band SSHH model defined on a square 
lattice, sketched in the lower right of Fig.~\ref{fig:Sallvsgholstfig1}, using determinant
quantum Monte Carlo (DQMC). Our main result is the low-temperature phase diagram 
shown in Fig.~\ref{fig:Sallvsgholstfig1}. It exhibits transitions between ${\bf Q} = (\pi/a,\pi/a)$ 
CDW, BOW, and AFM orders depending on the relative strengths of the SSH and Holstein couplings, 
$g_\mathrm{ssh}$ and $g_\mathrm{hol}$. In what follows, we will discuss the various numerical 
measurements leading to this phase diagram. 
Our results highlight the rich physics contained in $e$-ph models when one allows for 
coupling to multiple phonon branches, which may be relevant for understanding different 
classes of materials like the transition metal oxides.\\  

\noindent{\it Model \& Methods} --- 
We will compute the equilibrium properties of the `bond' SSHH model, where the in-plane 
phonon modes live on the links connecting pairs of sites~\cite{sengupta03peierls} and the
 Holstein modes live on the sites themselves. We consider a 2D square lattice with $N=L^2$
 sites, where $L$ is the linear size of the lattice. The Hamiltonian is 
\begin{align} \nonumber
    \mathcal{H} =& - t \sum_{\langle i, j \rangle, \sigma} 
\left( 1 - \alpha_{\rm ssh} \hat{X}_{ij} \right)
  \left( \hat{c}^{\dagger}_{i \sigma} \hat{c}^{\phantom\dagger}_{j \sigma}
 + {\mathrm{h.c.}} \right)
  - \mu \sum_{i, \sigma} \hat{n}_{i \sigma} \\\label{eq:ham}
                 & + \sum_{\langle i, j\rangle} \left(\frac{\hat{P}_{ij}^2}{2 M_\mathrm{ssh}} +
  \frac{M_\mathrm{ssh}}{2} \omega_\mathrm{ssh}^2 \hat{X}_{ij}^2 \right)  \\\nonumber
      & + \alpha_{\rm hol} \sum_{i,\sigma} \hat Y_i \hat{n}_{i\sigma} 
                  + \sum_{i} \left(\frac{\hat{\Pi}_{i}^2}{2M_\mathrm{hol}} +
  \frac{M_\mathrm{hol}}{2} \omega_\mathrm{hol}^2 \hat{Y}_{i}^2 \right). 
\end{align}
Here, $\hat{c}^{\dagger}_{j \sigma}$ $(\hat{c}^{\phantom\dagger}_{j \sigma})$
are fermionic creation (destruction) operators at site $j$ and with spin $\sigma$, $t$ is the 
nearest-neighbor hopping integral, and $\mu$ is the chemical potential. The SSH $e$-ph coupling in units of energy $g_{\rm ssh}$ is defined as $g_{\rm ssh}=\alpha_{\rm ssh}\sqrt{\hbar/(2M_{\rm ssh}\omega_{\rm ssh})}$; the Holstein coupling $g_{\rm hol}$ can be defined similarly with its accompanying parameters. $\hat X_{ij}$ and 
$\hat Y_i$ are independent SSH and Holstein phonons defined on the bonds and sites, respectively.
 The associated momentum operators for these phonons are $\hat{P}_{ij}$ and $\hat{\Pi}_i$, respectively. 
The effective masses are taken to $M_\mathrm{ssh}=M_\mathrm{hol}=1$, 
and the SSH and Holstein branches are dispersionless with energies $\omega_\mathrm{ssh}$ and 
$\omega_\mathrm{hol}$, respectively. The model includes two types of $e$-ph interactions:
 an on-site Holstein coupling $g_{\rm hol}$ to the fermionic density, and the SSH coupling
 $g_{\rm ssh}$ to the intersite hopping.  Throughout, we set $t = 1$ as our unit of energy, 
and adopt $\omega_\mathrm{hol}=\omega_\mathrm{ssh}=1$. This choice fixes the adiabatic
 ratio $\omega_\mathrm{ssh}/E_\mathrm{F}=\frac{1}{4}$ ($E_\mathrm{F} = 4t$ is the Fermi 
energy for the half-filled band) for both sets of modes and facilitates direct comparisons of 
the coupling strengths $g_{\rm hol}$ and $g_{\rm ssh}$. Finally, we fix 
$\mu = -2g_{\rm hol}^2/\omega_\mathrm{hol}$ to maintain half-filling.

We study Eq.~\eqref{eq:ham} using DQMC~\cite{blankenbecler81,assaad02}. This method 
expresses the finite temperature SSHH partition function as a path integral over
the space and imaginary-time dependent phonon fields $X_{ij}(\tau), Y_i(\tau)$.
 The fermionic degrees of freedom, which appear quadratically in the
Hamiltonian, are traced out analytically, so that the weight for the
phonon paths is a product of a bosonic contribution originating in the pure phonon
terms of Eq.~(\ref{eq:ham}) and a product of fermionic determinants of each spin species.
 The configurations of these fields are then sampled stochastically through
a combination of local moves at a single space-imaginary time value of each field [$X_{ij}(\tau)$, $Y_i(\tau)$], and
global moves that change the field at all imaginary time values $\tau$ simultaneously at 
a single spatial site $i$ or bond~\cite{Johnston13}. (The latter effectively 
samples configurations that are strongly correlated in the imaginary time direction due
 to the kinetic energy terms $\hat P_{ij}^2$ and $\hat \Pi_i^2$
in the Hamiltonian.) 
The algorithm scales as the cube of the number of spatial sites, and
(roughly) linearly with inverse temperature $\beta$. 
Importantly, there is no sign problem in our model at any filling~\cite{loh90, troyer05, mondaini22} 
owing to the symmetric coupling of the up and down fermionic species to the phonons. 

All runs begin with a ``seed input," in which phonon variables have a structure similar to 
the expected BOW or CDW phase, based on the relative values of the $e$-ph couplings 
$g_{\rm hol}$ and $g_{\rm ssh}$.
This practice helps reduce the number of warm-up sweeps that are needed to reach thermal 
equilibrium. Our code then runs through 320 imaginary time slices of 
$\beta=16$ ($\Delta\tau=0.05$) for a total of around $10^5$ steps.

We characterize the SSHH model by measuring the equilibrium structure factors associated
 with the different types of order. The relevant real space correlation functions are
 $C_\alpha({\bf r}) = \sum_{\bf i} \langle O_{\alpha}({\bf r}+{\bf i})O_{\alpha}({\bf i}) \rangle$,
 where $O_{s}({\bf r})= n_{{\bf r},\uparrow}-n_{{\bf r},\downarrow}$, 
$O_{c}({\bf r})= n_{{\bf r},\uparrow}+n_{{\bf r},\downarrow}$,
 and $O_{bx}({\bf r})= (c^{\dagger}_{{\bf r}+\hat x,\sigma}  c^{\phantom{\dagger}}_{{\bf r },\sigma} + \rm h.c.)$
 for spin ($s$), charge ($c$), and bond kinetic energy in the $\hat x$ direction ($bx$, with an analogous
 definition of $C_{by}$), respectively. The corresponding structure factors are obtained by a Fourier transform
 $S_\alpha({\bf q}) = \frac{1}{N} \sum_{\bf r} e^{\mathrm{i}{\bf q}\cdot{\bf r}}C_{\alpha}({\bf r})$. 
 Note, our normalization is such that the structure factors are lattice size independent in high temperature phases (short-range spatial correlations), but grow linearly with $N$ with the onset of long-range order. We will also present results for other standard observables like the average electron and phonon energies, double occupancy, etc. Additional details can be found in the supplementary materials. \\

\begin{figure}[t!]
\centering
\hskip-0.20in
\includegraphics[width=\columnwidth]{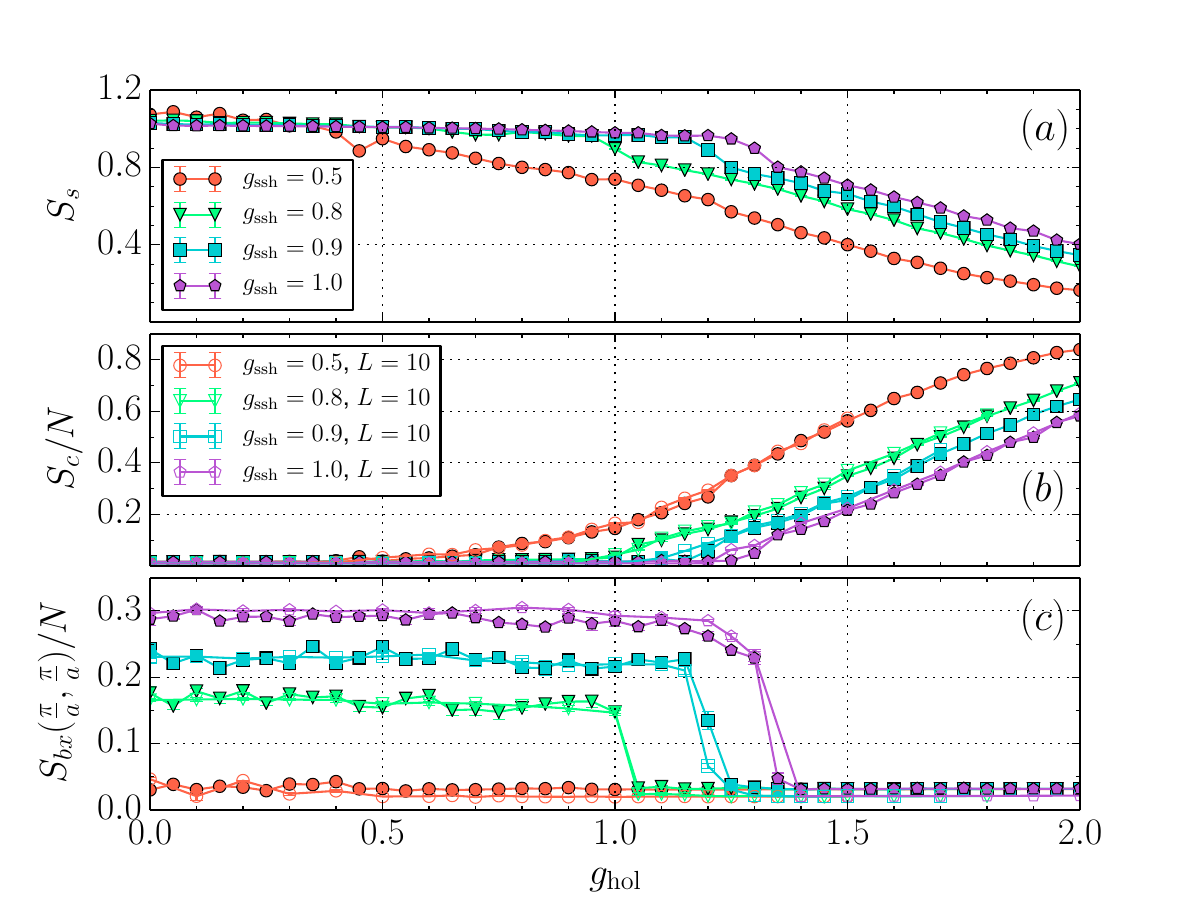} 
\caption{(a)
AF $S_{s}$, (b) CDW $S_{c}$, and (c) BOW $S_{bx}$ structure factors vs.~Holstein 
coupling $g_{\rm hol}$ at inverse temperature $\beta=16$, and half-filling for $L=8$
 (solid symbols) and $L=10$ (open symbols).  
An abrupt collapse of bond order occurs as $g_{\rm hol}$ grows for SSH couplings 
above the critical value, accompanied by a decrease in AF order, and a rise in CDW order.  Error bars are present, but smaller than marker size.
}
\label{fig:Scorrvsgholst} 
\end{figure}

\noindent{\it Results} --- To determine the phase boundaries shown in Fig.~\ref{fig:Sallvsgholstfig1}, 
we measured the evolution of the relevant structure factors in the $(g_\mathrm{hol},g_\mathrm{ssh})$ plane. 
Fig.~\ref{fig:Scorrvsgholst} plots results for 
the AFM [$S_s$, Fig.~\ref{fig:Scorrvsgholst}(a)], BOW [$S_{bx}$, Fig.~\ref{fig:Scorrvsgholst}(b)], 
and CDW [$S_{c}$, Fig.~\ref{fig:Scorrvsgholst}(c)] structure factors for a family of fixed $g_{\rm ssh}$
 while sweeping $g_{\rm hol}$ from weak to strong coupling. 
These results were obtained at a fixed inverse temperature $T = t/16$, which is low enough that the structure factors
reflect the ground state properties of the system for our parameters, i.e.,~the correlation length of the order in question exceeds the linear lattice size.

Focusing first on small $g_{\rm hol}$, we find that the model is dominated by BOW order for $g_{\rm ssh} > g_{\rm ssh,c} \approx 0.7$. This is reflected in the large value of $S_{bx}$ at small $g_{\rm hol}$, which grows in proportion to lattice size so that $S_{bx}/N$ is independent of $N$, indicative of long-range bond order.

\begin{figure}[t!]
\centering
\includegraphics[width=\columnwidth]{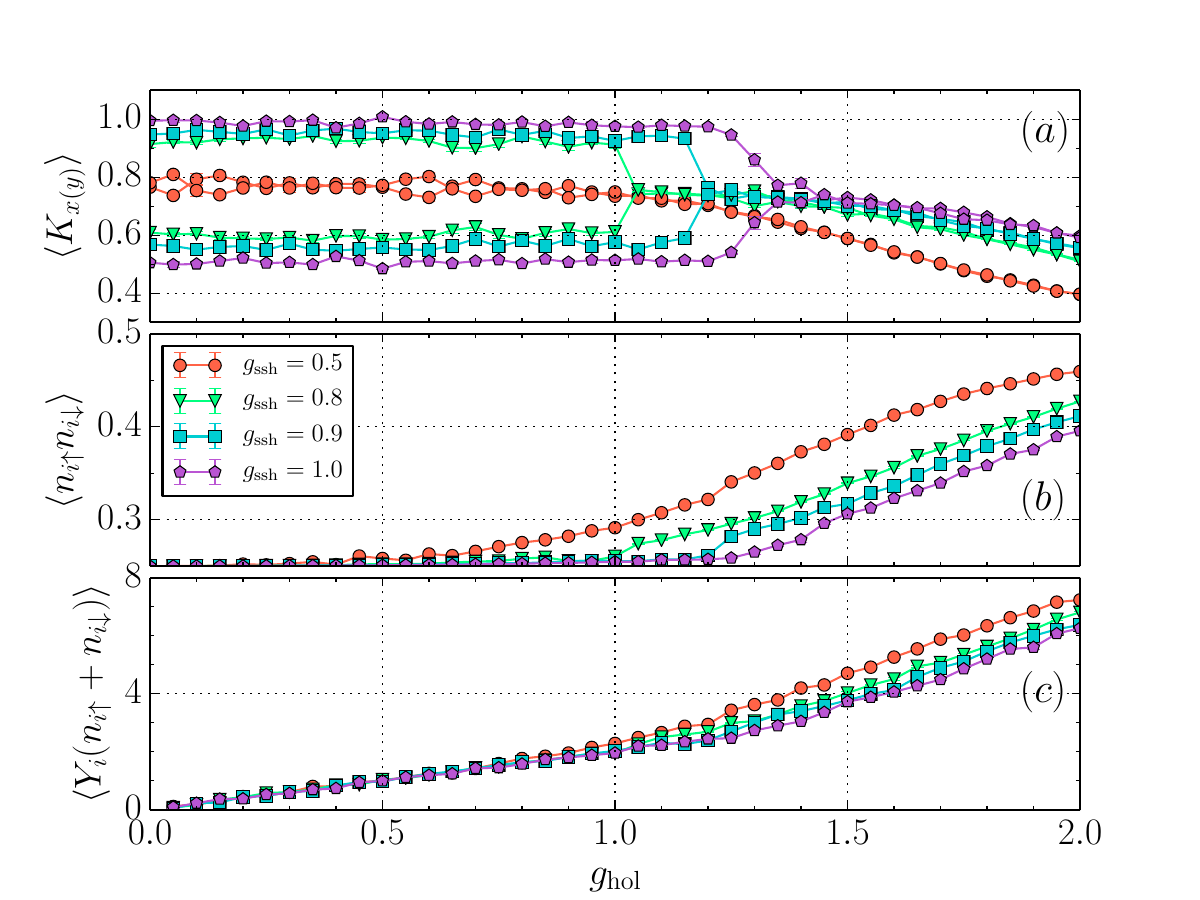} 
\caption{(a) 
$x$/$y$ kinetic energy, (b) double occupancy, and (c) Holstein electron-phonon coupling on an $L=8$ system at half-filling. 
Electronic kinetic energy and double occupancy are close to constant in the 
low $g_{\rm hol}$ regime where AF order is expected for $g_{\rm ssh} \lesssim 0.7$ and BOW are expected for $g_{\rm ssh} \gtrsim 0.7$.
}
\label{fig:Observvsgholstfig1} 
\end{figure}

As $g_{\rm hol}$ increases, $S_{bx}$ undergoes an abrupt collapse, suggesting a first order
 phase transition. The value of the critical $g_{\rm hol,c}$ decreases as $g_{\rm ssh}$ goes down, 
as does the amplitude of $S_{bx}$. At the same time, the strength of the CDW begins to increase
 continuously with $g_{\rm hol}$ once the BOW order has collapsed, as shown in 
Fig.~\ref{fig:Scorrvsgholst}(b). With the formation of CDW correlations, there is also 
a concomitant drop in the electron kinetic energy [Fig.~\ref{fig:Observvsgholstfig1}(a)], 
increase in the double occupations [Fig.~\ref{fig:Observvsgholstfig1}(b)], 
and a change in slope of the $e$-ph 
Holstein interaction energy [Fig.~\ref{fig:Observvsgholstfig1}(c)]. 
These results show that hopping is rapidly phased out in favor of ordered double occupancy as the Holstein coupling begins to dominate. This behavior is very similar to the formation of CDW order in the pure 2D Holstein model \cite{scalettar89,marsiglio90,vekic1992,freericks93,weber2018,costa2018,hohenadler19,zhang2019,cohen2019,chen2019,feng2020,nosarzewski21,bradley2021,araujo22}. 

These $g_{\rm ssh} \gtrsim 0.7$ results demonstrate the presence of a first order transition between a BOW and a CDW 
phase as a function of $g_{\rm hol}$ at large $g_{\rm ssh}$. However, the situation is 
quite different at lower values of $g_{\rm ssh}$. In the pure bond-SSH model, there is a 
phase transition from AF order to BOW as the value of $g_{\rm ssh}$ increases~\cite{feng21}. 
AF is stabilized in this model because the electrons can only simultaneously tunnel on a 
modulated bond if they are of opposite spin \cite{feng21}. This mechanism is in contrast to 
the AF order found in the Hubbard model, in which on-site repulsion $U$ penalizes double 
ccupancy while favoring the residual exchange interaction arranging sites antiferromagnetically. 

The AF order for small $g_{\rm ssh}$ in the pure SSH model persists
in the SSHH model for small $g_{\rm hol}$. 
For example, for $g_{\rm ssh}=0.5$ and small $g_{\rm hol}$ we remain in a state characterized 
by a higher magnitude of electronic kinetic energy accompanied by a weakly enhanced spin correlation 
$S_s$ [Fig.~\ref{fig:Scorrvsgholst}(a)]. In this case, the AF correlations are weak because
 the bonds are only weakly disturbed at small $g_{\rm ssh}$. Upon increasing either $g_{\rm ssh}$ 
or $g_{hol}$, the AF state gives way to BOW or CDW order.  The CDW phase boundary has a change
 in slope at $g_{\rm{hol}} \sim 1$.
This corresponds well to where the extrapolation of the AFM-BOW boundary intersects the CDW 
phase, demonstrating further consistency between the different order parameter measurements.

\begin{figure}[t!]
\centering
\includegraphics[width=\columnwidth]{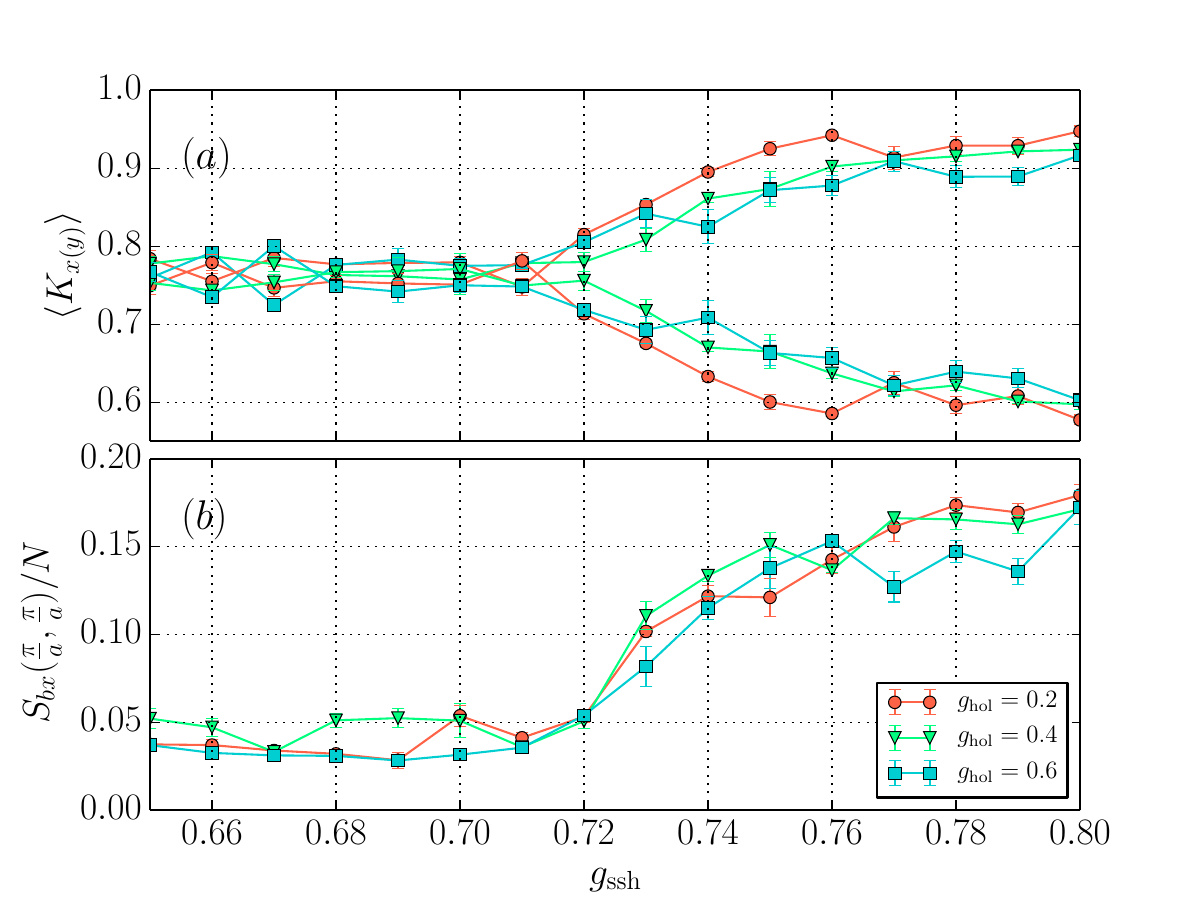} 
\caption{(a) $x$/$y$ kinetic energy and (b) CDW $S_{c}$ vs.~$g_{\rm ssh}$.  
Antiferromagnetic order feeds into BOW formation, transitioning at around $g_{\rm ssh,c}=0.71-0.72$.}
\label{fig:vertcuts1} 
\end{figure}

Fig.~\ref{fig:Observvsgholstfig1} plots several other relevant observables for the same parameter sets. 
The BOW order breaks the $x$/$y$ symmetry on a square lattice, since alternating large and
 small kinetic energy bonds select one of these axes along which to align. 
This phenomenon is evidenced in Fig.~\ref{fig:Observvsgholstfig1}(a). For $g_{\rm ssh}\ge 0.8$
 and small $g_{\rm hol}$, the system is in a BOW state and 
the kinetic energies along the $x$ and $y$ directions have two distinct values (differing by up 
to a factor of three). Increasing $g_{\rm hol}$ causes this bifurcation to collapse as the system 
transitions from the BOW phase to the CDW phase. For $g_{\rm ssh}=0.5$, there is no BOW at any 
$g_{\rm hol}$ and the $x$ and $y$ kinetic energies are equal across this cut through the phase diagram.

Having described the phase transitions which occur at fixed $g_{\rm ssh}$ with increasing $g_{\rm hol}$
(horizontal sweeps in Fig.~\ref{fig:Sallvsgholstfig1})
we can complete our picture of the phase diagram by studying 
vertical sweeps in Fig.~\ref{fig:Sallvsgholstfig1}, i.e.~trajectories 
at fixed $g_{\rm hol}$ with increasing $g_{\rm ssh}$.
These are given in Fig.~\ref{fig:vertcuts1}.
The expected transition into a BOW phase is confirmed by a sharp rise
in $S_{bx}$ 
[Fig.~\ref{fig:vertcuts1}(b)]
and a kinetic energy bifurcation
[Fig.~\ref{fig:vertcuts1}(a)].
The critical value of $g_{\rm ssh}$ is almost independent
of $g_{\rm hol}$, reflecting the nearly horizontal AF-BOW phase boundary of
Fig.~\ref{fig:Sallvsgholstfig1}. \\

\noindent{\it Conclusions} --- 
We have mapped the low temperature phase diagram of the 2D SSHH model at 
half-filling in the ($g_{\rm hol}, g_{ssh}$) plane. Starting in an AF phase found at weak 
couplings, we found that increasing $g_{\rm ssh}$ up to a critical value results in $x$/$y$ 
symmetry breaking as the intersite hopping modulates into a regular pattern, creating a BOW.
 For every value of $g_{\rm ssh}$, however, a critical value of $g_{\rm hol}$ exists that 
destroys the BOW and replaces it with a CDW.  We find  that the enhancement of BOW at higher 
$g_{\rm ssh}$ delays the CDW onset
(i.e.~requiring larger $g_{\rm hol}$); this behavior is due to the direct competition between the phases,
 as BOW favors quantum fluctuations and CDW prefers electron localization.
In many models with competing interactions, e.g.~the 1D extended Hubbard 
Hamiltonian~\cite{baeriswyl1985theoretical,emery1979highly,nakamura2000tricritical}, 
the transition between different phases changes from second order at weak coupling to 
first order at strong coupling.
We do not see firm evidence for such a tricritical point here.  However, it is possible
 we are not yet at strong enough coupling.  The largest value of the effective (attractive)
 $U_{\rm eff}=-g_{\rm hol}^2/\omega_\mathrm{hol}^2$
in our phase diagram of Fig.~\ref{fig:Sallvsgholstfig1} 
is $|U_{\rm eff}| \sim 2t$.  The strong coupling (first order) character in the 1D 
extended Hubbard model
occurs only beyond $(U_t,V_t)=(5.89t,3.10t)$~\cite{ejima2007phase}, and would likely 
require even larger values in the 2D
geometry studied here.

Our work parallels exploration of the interplay 
of individual forms of $e$-ph interaction with 
on-site electron-electron interactions (a Hubbard $U$)~\cite{Johnston13, weber2018, Costa20}.
In the case of the Hubbard-SSH Hamiltonian at half-filling  there is no
fermion sign problem~\cite{TanjaroonLy2023comparative,loh90,troyer05,mondaini22}
and the low temperature antiferromagnetic to BOW phase can be
effectively characterized~\cite{feng21}.  As this work was done entirely for $\beta=16$,
 a fruitful investigation would be to characterize the transition temperatures between 
the AF and BOW transitions in the presence of Holstein phonons. \\

\noindent{\it Acknowledgements} --- 
This work was supported by the  U.S.~Department of Energy, Office of Science, Office of 
Basic Energy Sciences, under Award Number DE-SC0022311. The Flatiron Institute is a division
 of the Simons Foundation.

\bibliography{sshholstein.bib}

\end{document}


\title{Supplementary Material for ``Magnetic, charge, and bond order in the two-dimensional Su-Schrieffer-Heeger-Holstein model''}
\author{Max Casebolt}
\affiliation{Department of Physics, University of California, Davis, CA 95616,USA}
\author{Chunhan Feng}
\affiliation{Center for Computational Quantum Physics, Flatiron Institute, 162 5th Avenue, New York, NY, USA}
\affiliation{Department of Physics, University of California, Davis, CA 95616,USA}
\author{Richard T. Scalettar}
\affiliation{Department of Physics, University of California, Davis, CA 95616,USA}
\author{Steven Johnston}
\affiliation{Department of Physics and Astronomy, The University of Tennessee, Knoxville, TN 37996, USA}
\affiliation{Institute for Advanced Materials and Manufacturing, University of Tennessee, Knoxville, TN 37996, USA\looseness=-1}
\author{G. G. Batrouni}
\affiliation{Centre for Quantum Technologies, National University of
  Singapore, 2 Science Drive 3, 117542 Singapore}
\affiliation{Universit\'e C\^ote d'Azur, CNRS, Institut de Physique de Nice (INPHYNI), 06000 Nice, France}

\newcommand\tlc[1]{\texorpdfstring{\lowercase{#1}}{#1}}
\renewcommand{\thetable}{S\arabic{table}}  
\renewcommand{\thefigure}{S\arabic{figure}}

\begin{abstract}
\end{abstract} 

\maketitle

Here we provide results for some additional measurements in support of our conclusions. They include filling versus chemical potential (Sec.~A1) and superconducting correlations (Sec.~A2), followed by data verifying that $\beta=16/t$ is sufficiently
large to be representative of the ground state (Sec.~A3).

\section{A1.~Average Filling vs. Chemical Potential}
\begin{figure}[h]
\centering
\includegraphics[width=0.5\columnwidth]{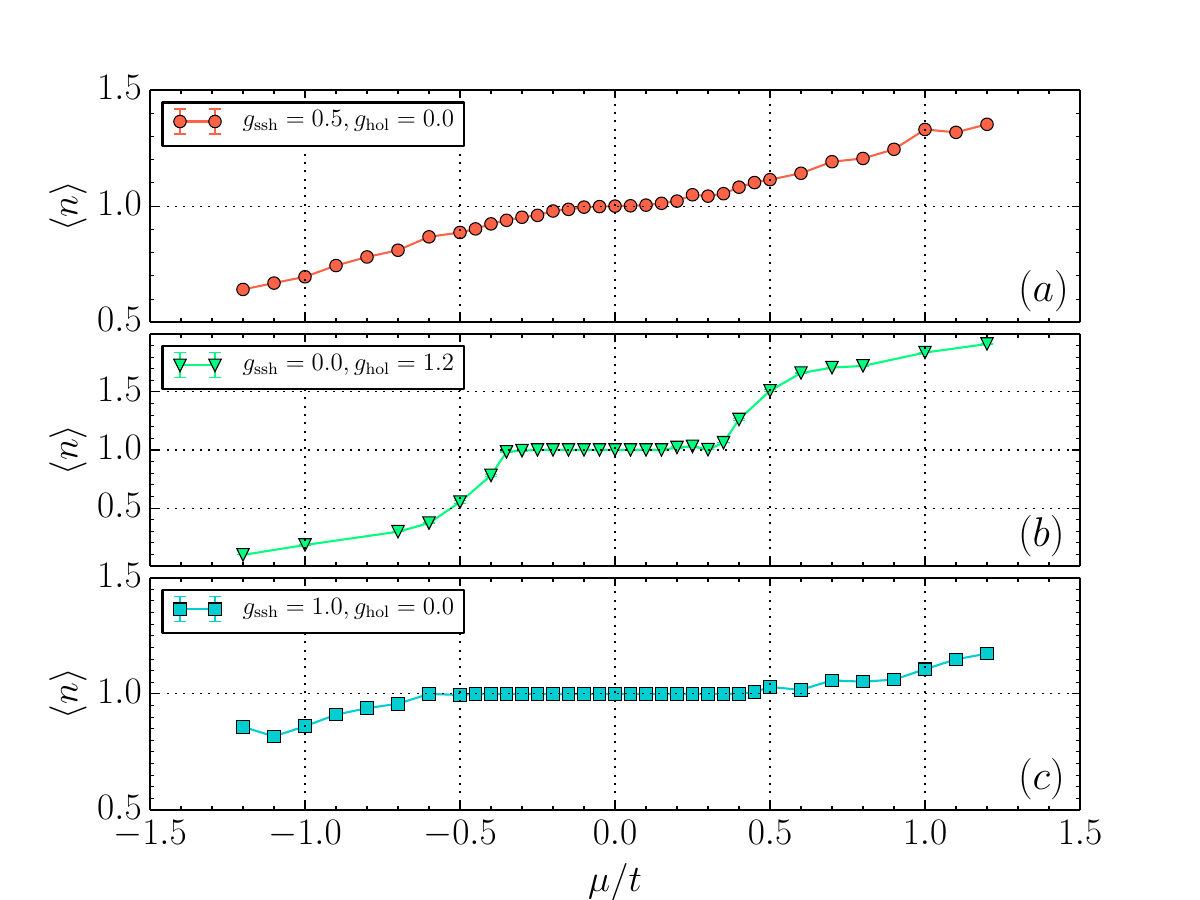} 
\caption{The average site density $\langle n \rangle$ 
as a function of the chemical potential $\mu$ at $\omega_{\rm ssh}=\omega_{\rm hol}=1$, recorded in 
three representative points ($g_{\rm ssh}$, $g_{\rm hol}$) of the phase diagram: (a) AFM, (b) CDW, and (c) BOW.  A shift in the half-filling value of $\mu$ is present via $-\frac{2g_{\rm hol}^2}{\omega_{\rm hol}}$, and these plots have been shifted by said amount to coincide with half-filling at $\mu=0$.
}

\label{fig:rhovsmufig1} 
\end{figure}

Figure~\ref{fig:rhovsmufig1} plots the average site density $\langle n \rangle = \frac{1}{N}\sum_{i,\sigma}\langle c^\dagger_{i,\sigma}c^{\phantom\dagger}_{i,\sigma}\rangle$ as a function of the chemical potential $\mu$ for three representative points in the  ($g_{\rm ssh}$, $g_{\rm hol}$) phase diagram. Specifically, curves are shown in the AFM (0.5, 0.0), CDW (0.0, 1.2), and BOW (1.0, 0.0) phases. All three curves show a plateau around half-filling, indicating that all three phases have a gap in their single-particle energy spectrum. In other words, they are insulating. The plateaus of CDW/BOW order [Fig.~\ref{fig:rhovsmufig1}(b,c)] are notably longer than that of AFM order. This is due to the types of symmetries being broken in such regions: continuous for AFM order, and discrete for CDW/BOW order.  One may also notice the sharper approach to the CDW plateau in Fig.~\ref{fig:rhovsmufig1}(b), a feature that has also been reported in Fig.~7 of Ref.~\cite{Bradley2021}.

\newpage

\section{A2.~Superconducting correlations}

\begin{figure}[h]
\centering
\includegraphics[width=0.5\columnwidth]{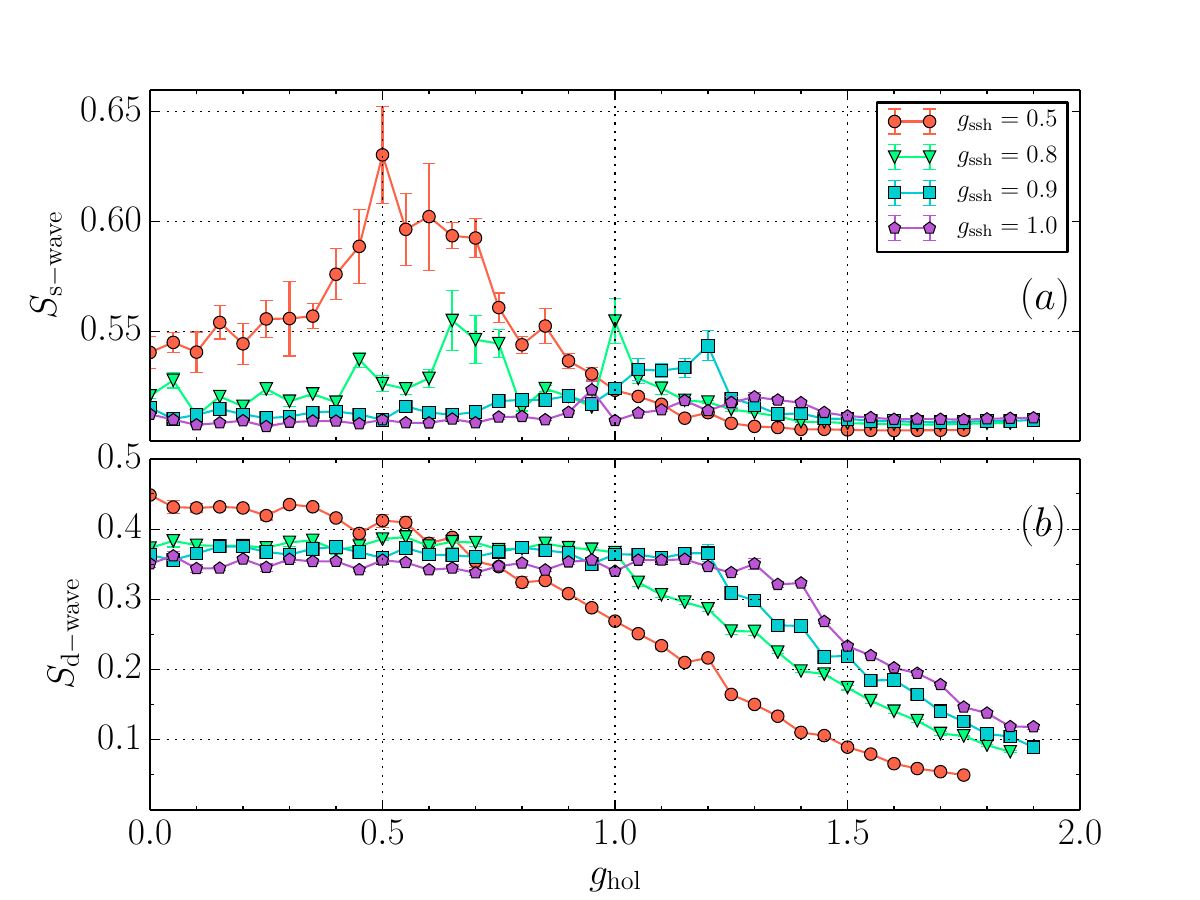} 
\caption{
Pairing structure factor $S_{\text{s-wave}}$ (a) and $S_{\text{d-wave}}$ (b)
as functions of Holstein coupling $g_{\rm hol}$ for four fixed values of $g_{\rm ssh}$.
The $s$-wave structure factor has a peak at the AF-CDW boundary, though the value is too small to be associated 
with long range order.
The $d$-wave structure factor is everywhere small (short ranged in real space), but it is largest
in the AF phase.  This parallels the known connection between spin density wave order and $d$-wave pairing
in the Hubbard model.
}
\label{fig:SdandSsfactors} 
\end{figure}

Figure~\ref{fig:SdandSsfactors}a shows the $s$-wave pairing structure factor, $S_{s-{\rm wave}}$, at different fixed
$g_{\rm ssh}$ as a function of $g_{\rm hol}$.  The Holstein model itself has a CDW ground state 
at half-filling, $\langle n \rangle= 1$, and will support superconductivity when doped.  We interpret the peak in $S_{s-{\rm wave}}$ to be associated with release from
the suppression of the pairing associated with CDW formation when $g_{\rm ssh} \gtrsim g_{\rm hol}$.

In the Hubbard model $d$-wave superconductivity is associated with AF correlations, which are strongest near half-filling.
Is there a similar connection to the AF that arises due to an SSH phonon mode?
Fig.~\ref{fig:SdandSsfactors}(b) shows the pairing structure factor $S_{d-\rm{wave}}$ at different fixed
$g_{\rm ssh}$ and increasing $g_{\rm hol}$.  We observe that $S_d$ turns downward to lower values upon exiting the AF phase
to the CDW.  Likewise, comparing the plots for increasing $g_{\rm ssh}$ we see that $S_d$ falls as the BOW phase is approached.Thus there appears also to be a correlation between $d$-wave pairing and AF in the present model,
though it should be noted that $S_d$ is too small to be consistent with long range pairing order.

\section{A3.~Verification of ground state}

\begin{figure}[h]
\centering
\includegraphics[width=0.5\columnwidth]{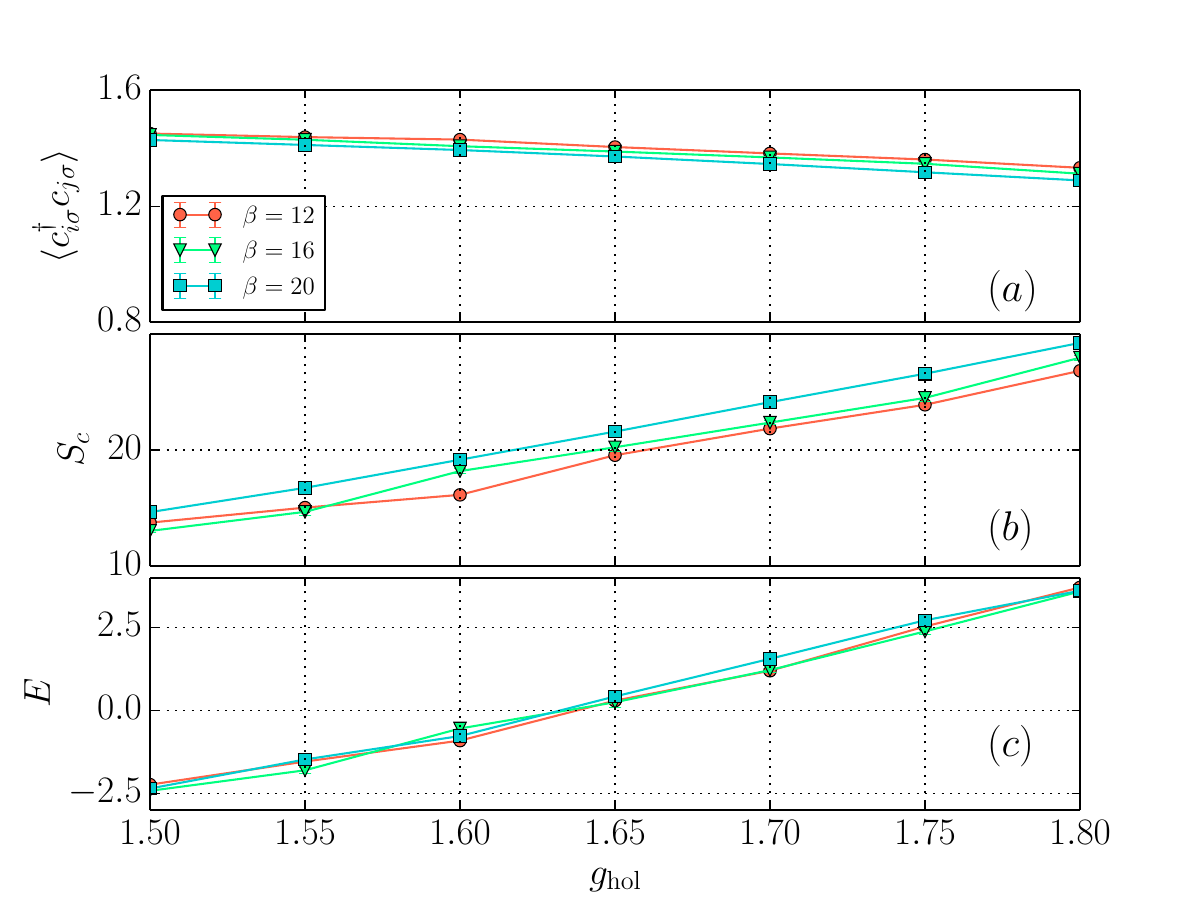} 
\caption{Kinetic energy (a), charge structure factor $S_{c}$ (b), and total energy (c) for $\beta=12/t, 16/t,$ and $20/t$.
The SSH coupling $g_{\rm ssh}=1$.
Since the data for $\beta=16/t$ and $\beta=20/t$ coincide, we conclude $\beta=16/t$ is sufficiently large to be sampling
the properties of the ground state.
}
\label{fig:systematics1} 
\end{figure}

Fig.~\ref{fig:systematics1} shows three measurements: kinetic energy, charge structure factor, and total energy,
at $\beta=12/t ,16/ t$ and $20/t$ for $g_{\rm ssh}=1.0$ and $1.5\leq g_{\rm hol}\leq 1.8$.  Results for $\beta= 16/t$
overlap well with lower temperature data $\beta/t=20$, to within statistical errors, indicating that
$\beta=16/t$ is sufficiently large to use throughout this work in sampling the $T=0$ phases.
We have verified the same is true throughout the phase diagram of Fig.~1 of the main paper.

\bibliography{sshholstein.bib}